\documentclass[showpacs,showkeys,amsmath,amssymb,preprint,nofootinbib,floats]{revtex4}
\usepackage{graphicx}
\usepackage{epsfig}
\usepackage{bm}
\usepackage{float}
\usepackage{fontenc}
\usepackage{graphicx}
\usepackage{bm}
\usepackage{amsmath}
\def\beq{\begin{eqnarray}}

\def\eeq{\end{eqnarray}}

\begin{document}

\title[Chaos]{Qualitative description of the universe in the interacting fluids scheme}

\author{Miguel Cruz$^{1,}$\footnote{E-mail: miguelcruz02@uv.mx}, Samuel Lepe$^{2,}$\footnote{E-mail: samuel.lepe@pucv.cl} and Gerardo Morales-Navarrete$^{1,}$\footnote{E-mail: moralesnavarretegerardo@gmail.com}}
\affiliation{$^{1}$Facultad de F\'\i sica, Universidad Veracruzana 91000, Xalapa, Veracruz, M\'exico \\ 
$^{2}$Instituto de F\'{\i}sica, Facultad de Ciencias, Pontificia Universidad Cat\'olica de Valpara\'\i so, Avenida Brasil 2950, Valpara\'\i so, Chile}

\date{\today}

\begin{abstract}
In this work we present a qualitative description of the evolution of a curved universe when we consider the interacting scheme for the constituents of the dark sector. The resulting dynamics can be modeled by a set of Lotka-Volterra type equations. For this model a future singularity is allowed, therefore the cyclic behavior for the energy interchange between the components of the universe is present only at some stage of the cosmic evolution. Due to the presence of the future singularity, the model exhibits global instability.
\end{abstract}



\maketitle
\section{Introduction}
\label{sec:intro}
The construction of a successful unified model within the standard cosmology for the description of the cosmic evolution is not yet achieved, i.e., for almost all the cosmological models the early time universe has not a smooth transition to the late time accelerating nature. In general, these two phases are studied separately and from different points of view. However, this does not mean that the search for a model of this type has not been done, it is worthy to mention that certain advances have been obtained when working beyond the standard cosmological model. In this sense, some approaches have more advantages than other.\\ 

A recurrent scenario of great interest is the so-called {\it interaction scheme}, in this framework two fluids are allowed to interact through a $Q$-function. This function determines the rate of energy transference between the fluids and it is chosen conveniently depending on the model \cite{q1, q2, q3, q4}. In a natural way this scheme has propitiated a description of the cosmic evolution from a more intuitive thermodynamic perspective, in this picture the changes in the sign of the $Q$-function could give some hints about the conditions that lead to thermodynamic equilibrium or possible phase transitions characterized by a change in the sign of the heat capacity \cite{phase1, phase}.\\ 

Moreover, the observational evidence for the accelerated expansion \cite{riess, perlmutter, tegmark, observations, observations1, des} requires a new component to drive such acceleration, often called {\it dark energy}. This new component has a negative pressure, but in order to explain the origin of the galaxies and their distribution, a component called {\it dark matter} must enter the game \cite{dm1, dm2}, with the characteristic of having constant temperature and pressure equal to zero. When the cosmic expansion is guided by the cosmological constant we have the well known $\Lambda$CDM model, but, as it is also known, some inconsistencies have been revealed for this model, where the cosmological constant value problem can be highlighted. See for instance the Ref. \cite{weinberg}, where was pointed out that there is a discrepancy between the value obtained from observations for the cosmological constant and the one obtained in particle physics. This has motivated the consideration of dynamical dark energy models. To sum up, the composition given by the dark energy - dark matter synergy\footnote{Or commonly termed {\it dark sector}.} it is necessary to explain the current status of the universe. In Ref. \cite{q5} can be found a complete review on the dark energy - dark matter interaction. On the other hand, in Ref. \cite{carde} the dark energy - dark matter interaction is described with the use of observational data and thermodynamics conditions.\\

In the meantime the dark energy problem, as exotic as it sounds, is still the missing puzzle of the modern cosmology, despite the efforts done by the community, it remains as an open subject of investigation. With this preamble, we must emphasize that in this work we will consider the dark energy - dark matter interaction together with the holographic approach for the description of the dark energy. The holographic framework is based on the existing relation between the ultraviolet and infrared cuts-off imposed by the formation of a black hole \cite{bhf}, then the density caused by the ultraviolet cut-off enclosed in a region of dimension, $L$, it should be similar to the mass of a black hole of the same size, the Hubble scale was considered as a first option for the characteristic length, $L$, resulting that with this choice the corresponding density is comparable with the value of the dark energy at present time \cite{l1, li, l3, l4}. In recent times, the holographic approach to the dark energy problem has been used recurrently since it has been found that it can provide some solutions to problems such as the cosmological coincidence problem, it yields an unified model for the early and late universe. Additionally, a description for the dark energy from a quantum point of view can be established and for some specific characteristic lengths, the phantom scenario can arise naturally \cite{cruz, l5, l6, l7, l8, l9}. For a complete review on the dark energy problem and holography, the Ref. \cite{holrev} can be seen.\\

Following the line of reasoning of Ref. \cite{aydiner}, it is proposed that in order to try to provide a more realistic model of the universe, the dark sector can interact with other components of the universe such as ordinary matter, this consideration leads to a dynamics that can be described through a set of Lotka-Volterra type equations when the interaction term $Q$ is chosen appropriately. As relevant result, the energy interchange in the dark sector exhibits a cyclic nature and the universe presents an attractor behavior when more components are considered. In this work we will employ the aforementioned proposal for a holographic dark energy model studied in Ref. \cite{cruz} within the interaction framework between dark energy and dark matter, as it was shown in the previous reference, a future singularity can take place only for a curved universe under this picture, this model allowed the construction of the interaction $Q$-term, therefore in this approach the election of this term it is not arbitrary. In this description the attractor behavior is not longer present and the cyclic behavior in the dark sector is only transient, as we will see later. It is worthy to mention that we will focus only on the interaction within the dark sector since its coupling with other components it is not well understood, in fact, the dark energy - dark matter interaction is expected to be small in order to be in agreement with the concordance model\footnote{The $\Lambda$CDM model is generally the best choice to fit observational data, the dark energy is modeled by the cosmological constant and in consequence is not interacting.}. On the other hand, the coupling of dark energy with baryons is probably negligible and additionally the coupling with radiation is much more difficult since the trajectories of photons would not follow null geodesics \cite{holrev, bolotin}.\\

This work is organized as follows: In Section \ref{sec:HQterm} we give a general description of the holographic model for a curved universe in the interacting scheme. We provide some details in the construction of the $Q$-term. At the end of this section we re-write the quantities found previously but now considering the singular behavior admitted by the model. In Section \ref{sec:chaotic} we show that the interaction term (which carries the singular nature exhibited by the model) can be written as a product of the densities of the dark sector components; this allows to describe the dynamics of the model as a set of Lotka-Volterra type equations. We show that the cyclic behavior in the dark sector is transient since eventually the components diverge. In Section \ref{sec:final} we present the final comments of our work.   
     
\section{Holographic $Q$-term in curved spacetime}
\label{sec:HQterm}
In this section we will provide a general description of the results obtained in Ref. \cite{cruz}, where was found that under the consideration of the interacting scheme for the dark matter - dark energy components of the universe, it is possible to construct the $Q$-term that mediates the interaction between the aforementioned components, when an appropriate holographic cut-off for the dark energy density is chosen and spatial curvature is included. For a FLRW curved spacetime the Friedmann constraint takes the following form
\begin{equation}
E^{2}(z) = \frac{1}{3H^{2}_{0}}\left(\rho_{DE}(z)+\rho_{DM}(z) \right) + \Omega_{k}(z),
\label{eq:friedmann}
\end{equation}
by $E(z)$ we will denote the normalized Hubble parameter, i.e., $E(z):= H(z)/H_{0}$ and will be given as a function of the redshift, $z$, we must take into account that the redshift and the scale factor, which is generally denoted by $a$, can be related each other by virtue of the standard definition, $1+z = a_{0}/a$. Throughout this work, the subscript $0$ will mean evaluation of the cosmological quantities at present time ($z=0$ or $a=a_{0}$). $\rho_{DE}$ and $\rho_{DM}$ stand for the dark energy and dark matter densities, respectively. Therefore, the third term on the r.h.s. of Eq. (\ref{eq:friedmann}) represents  the curvature parameter which is defined as, $\Omega_{k}(z) := \Omega_{k}(0)(1+z)^{2}$, where $\Omega_{k}(0)$ is a constant defined as follows, $\Omega_{k}(0) = -k/a^{2}_{0}H^{2}_{0}$. The parameter $k$ represents the topology of the spacetime and can take the values $\pm 1, 0$. In general, when interacting fluids are considered, the function that measures the energy transference between the fluids or simply, $Q$-term, is introduced in the continuity equations for the energy densities as written below
\begin{align}
& \rho'_{DE} - 3\left(\frac{1+\omega_{DE}}{1+z} \right)\rho_{DE} = \frac{Q}{H_{0}E(z)(1+z)}, \label{eq:cont1}\\
& \rho'_{DM} - \left(\frac{3}{1+z} \right)\rho_{DM} = -\frac{Q}{H_{0}E(z)(1+z)}, 
\label{eq:cont2}
\end{align}  
the prime stands for redshift derivative and $\omega$ it is the well known parameter state, which relates the pressure of the fluid with its energy density. For the dark matter sector we have considered, $\omega_{DM} = 0$. If we consider the Eqs. (\ref{eq:friedmann}), (\ref{eq:cont1}) and (\ref{eq:cont2}), we obtain the following expression 
\begin{equation}
1+\frac{\omega_{DE}(z)}{1+r(z)} = \frac{2}{3}\left(\frac{1}{2}(1+z)\frac{d \ln E^{2}(z)}{dz}-\Omega_{k}(0)\left(\frac{1+z}{E(z)}\right)^{2}\right)\left[1-\Omega_{k}(0)\left(\frac{1+z}{E(z)}\right)^{2}\right]^{-1},
\label{eq:omega0}
\end{equation}
where $r(z)$ is the coincidence parameter defined as $r(z) := \rho_{DM}(z)/\rho_{DE}(z)$. From now on, by means of the holographic principle, the energy density $\rho_{DE}$ will be given in terms of the Hubble scale \cite{li}
\begin{equation}
\rho_{DE} = 3c^{2}H^{2}_{0}E^{2}(z),
\label{eq:holo}
\end{equation}
where $c^{2}$ is a convenient constant that enters in the conventional expression for the holographic dark energy (\ref{eq:holo}) and $0 < c^{2} < 1$. Note that once we assume a specific holographic form for the dark energy density and using the Friedmann constraint (\ref{eq:friedmann}), after a straightforward calculation we can obtain the corresponding expression for the dark matter energy density, $\rho_{DM}$, yielding
\begin{equation}
\rho_{DM} = 3H^{2}_{0}E^{2}(z)\left[1-c^{2}-\Omega_{k}(0)\left(\frac{1+z}{E(z)}\right)^{2}\right].
\label{eq:holodm}
\end{equation}  
If we insert the previous result in the continuity equation (\ref{eq:cont2}), one gets
\begin{equation}
(1+z)\frac{d \ln E^{2}(z)}{dz} = 3-\frac{1}{1-c^{2}}\left[\Omega_{k}(0)\left(\frac{1+z}{E(z)}\right)^{2}+\frac{Q}{3H^{3}_{0}E^{3}(z)}\right].
\label{eq:cont3}
\end{equation}
Using the previous result in Eq. (\ref{eq:omega0}) we have
\begin{eqnarray}
\frac{Q(z)}{9(1-c^{2})H^{3}_{0}E^{3}(z)} = 1-\Omega_{k}(0)\left(\frac{1+z}{E(z)}\right)^{2}\left(\frac{3-2c^{2}}{3(1-c^{2})}\right)&-&\left(1+\frac{\omega_{DE}(z)}{1+r(z)}\right)\times \nonumber \\ &\times & \left[1-\Omega_{k}(0)\left(\frac{1+z}{E(z)}\right)^{2}\right].
\label{eq:interact}
\end{eqnarray}
It is important to point out that what was expressed in the previous equation for the interaction term emerges as a construction based on the dynamics of the model, contrary to what usually happens in the interaction scheme, we do not consider an Ansatz for the $Q$-term. In general, the election of the $Q$-term is based on its viability to describe the interaction in concordance to the observational data \cite{q1, q2, q3, q4, q5, q6, q7}. If we consider the definition of the deceleration parameter given as a function of the redshift,  $1+q(z) = (1+z)(d\ln E(z)/dz)$, therefore from the Eq. (\ref{eq:cont3}), the $Q$-term can be related to the deceleration parameter as follows
\begin{equation}
q(z) = \frac{1}{2}\left(1-\frac{1}{1-c^{2}}\left[\Omega_{k}(0)\left(\frac{1+z}{E(z)}\right)^{2}+\frac{Q}{3H^{3}_{0}E^{3}(z)}\right]\right).
\label{eq:decel}
\end{equation}
At present time both equations (\ref{eq:interact}) and (\ref{eq:decel}), take constant values. Without loss of generality we can have at present time $Q_{0} > 0$ always that $q_{0} < 1/2$. On the other hand, for $q(z) > 1/2$ we will have $Q(z) < -3H_{0}^{3}E(z)\Omega_{k}(0)(1+z)^{2}$, note that the change in the sign of the $Q$-term depends only on the value of the curvature parameter. The changes in the sign of the interaction term have raised interest from a thermodynamical point of view since it was found that such changes can be associated with the existence of phase transitions during the cosmic evolution \cite{phase1, phase2}. 
    
\subsection{Singular behavior}
\label{sec:singular}
As commented previously, using the holographic cut-off given in Eq. (\ref{eq:holo}) for the dark energy density, $\rho_{DE}$, together with the Friedmann constraint (\ref{eq:friedmann}), we can obtain the corresponding expression for $\rho_{DM}$ given in (\ref{eq:holodm}), therefore the coincidence parameter has the form
\begin{equation}
r(z) = \frac{1}{c^{2}}\left[1-c^{2}-\Omega_{k}(0)\left(\frac{1+z}{E(z)}\right)^{2}\right],
\label{eq:coincidence}
\end{equation}
from the previous expression we can write
\begin{equation}
E^{2}(z) = -\frac{\Omega_{k}(0)(1+z)^{2}}{c^{2}\left(r(z)-r_{c}\right)},
\label{eq:singular} 
\end{equation}
where we have defined the constant quantity, $r_{c} := (1-c^{2})/c^{2}$. For $r(z) = r_{c}$, we have a singularity for the normalized Hubble parameter, as pointed out in Ref. \cite{cruz}, the singularity admitted by the model it is of Type III. A complete classification and description of the future singularities that can appear in a dark energy model can be found in Ref. \cite{odintclass}. By considering a Chevallier-Polarsky-Linder type parametrization for the coincidence parameter, i.e., $r(z) = r_{0}+\epsilon_{0}[z/(1+z)]$, where $\epsilon_{0} > 0$ and solving the singularity condition $r(z) = r_{c}$, we obtain
\begin{equation}
z_{s} = -\frac{r_{0}-r_{c}}{\epsilon_{0}\left(1+(r_{0}-r_{c})/\epsilon_{0}\right)},
\label{eq:zsing}
\end{equation}
which is the value of the redshift at which the singularity takes place, from previous equation we have the condition, $-1 < z_{s} < 0$. If we use the previous results in Eq. (\ref{eq:singular}) we can write
\begin{equation}
E^{2}(z) = -\eta \Omega_{k}(0)\frac{(1+z)^{3}}{z-z_{s}},
\label{eq:normalized}
\end{equation}
for simplicity in the notation we have defined the following constant, $\eta := (1+z_{s})/c^{2}\epsilon_{0}$. Finally, by introducing the function $\theta(z) := (1+z)/(z-z_{s})$ and by direct substitution of Eq. (\ref{eq:normalized}) into Eq. (\ref{eq:omega0}), we can write
\begin{equation}
1+\frac{\omega_{DE}(z)}{1+r(z)} = \frac{2-\eta \theta(z)(\theta(z)-3)}{3\left[1+\eta \theta(z) \right]},
\label{eq:limit}
\end{equation}
and following a similar procedure, the Eq. (\ref{eq:interact}) takes the form
\begin{equation}
\frac{Q(z)}{3H^{3}_{0}} = -\Omega_{k}(0)\sqrt{-\Omega_{k}(0)\eta \theta(z)}(1+z)^{3}\left[1+\eta \theta^{2}(z)\left(1-c^{2}\right) \right],
\label{eq:Qtheta}
\end{equation}   
which exhibits a singular behavior when $z = z_{s}$. As stated in Ref. \cite{q6}, any future singularity induced in the interacting scheme can be mapped into a singular behavior of the interaction $Q$-term.
 
\section{Interacting description}
\label{sec:chaotic}
In this section we implement a qualitative description of an universe in which the interaction between some of its components is allowed, this is done following the line of reasoning of Ref. \cite{aydiner}, where it was found that the interaction between the constituents of the universe can be modelled through some Lotka-Volterra type equations. In order to describe an universe in which a future singularity can occur, we will consider as interaction $Q$-term the one given in Eq. (\ref{eq:Qtheta}). If we consider the expression for the normalized Hubble parameter given in Eq. (\ref{eq:normalized}) together with the $\theta(z)$ function, the densities for the dark energy and dark matter can be written as follows
\begin{align}
& \rho_{DE} = -3c^2\eta \theta(z) H_{0}^{2} \Omega_{k}(0) (1+z)^{2},\\
& \rho_{DM} = -3H_{0}^{2} \eta \Omega_{k}(0) \theta(z)(1+z)^2 \left( 1-c^{2}+ \frac{1}{\eta \theta(z)} \right),
\end{align}
where the singularity condition can be found in the $\theta(z)$ function. Using the previous equations in the $Q$-term (\ref{eq:Qtheta}), one gets 
\begin{equation}
Q(z) = - \frac{1}{3c^{2}H_{0}} \frac{1}{\sqrt{-\Omega_{k}(0) \eta \theta(z)}}\frac{ 1+ \eta \theta^2(z) (1-c^2)}{ 1+ \eta \theta(z) (1-c^2)}\rho_{DM} \rho_{DE}.
\end{equation}
It is important to point out that the construction itself of the $Q$-term allows us to write it as a product of the dark energy and dark matter densities, in Ref. \cite{aydiner} this kind of $Q$-term is also used, but, it is given as an Ansatz. Therefore, from the continuity equations (\ref{eq:cont1}) and (\ref{eq:cont2}), we can write
\begin{align}
& \frac{d\rho_{DE}}{dz} = 3\left(\frac{1+\omega_{DE}}{1+z} \right)\rho_{DE} + \beta(z)\rho_{DM}\rho_{DE}, \label{eq:contrhos1}\\
& \frac{d\rho_{DM}}{dz} = \left(\frac{3}{1+z} \right)\rho_{DM} - \beta(z)\rho_{DM}\rho_{DE}, 
\label{eq:contrhos2}
\end{align}  
where we have defined the function 
\begin{equation}
\beta(z) = \frac{1}{3c^{2}H^{2}_{0}\eta \Omega_{k}(0) \theta(z)(1+z)^2}\frac{ 1+ \eta \theta^{2}(z) (1-c^{2})}{ 1+ \eta \theta(z) (1-c^{2})}.
\end{equation}
As can be seen from previous equation, the $\beta(z)$ function can change its sign and this strongly depends on the value of the curvature parameter. In general, we can see that $\beta(z) > 0$ since $z > z_{s}$. On the other hand, if we define the quantities $r_{1} := 3(1+\omega_{DE})$ and $r_{2} := 3$, we can see that $r_{1} < 0$ since the parameter state $\omega_{DE}$ can cross to the phantom region. In terms of these quantities, the continuity equations (\ref{eq:contrhos1}) and (\ref{eq:contrhos2}) can be expressed as follows
\begin{align}
& \frac{dx_{1}}{dz} = r_{1}x_{1}\left(\alpha(z)+\beta(z)x_{2}\right), \label{eq:contx1}\\
& \frac{dx_{2}}{dz} = r_{2}x_{2}\left(\alpha(z)-\beta(z)x_{1}\right), 
\label{eq:contx2}
\end{align}  
where $x_{1} := \rho_{DE}/r_{2}$, $x_{2}:=\rho_{DM}/r_{1}$ and $\alpha(z):= 1/(1+z)$. It is worthy to mention that previous equations have the structure of the Lotka-Volterra equations, but, we can say that in our case we have a generalization of these since the functions $\alpha(z)$ and $\beta(z)$ appear as coefficients of the $x_{1,2}$ {\it species}. Besides, these equations can be obtained only because the form of the $Q$-term is given as a product of the dark energy and dark matter densities, as mentioned in Ref. \cite{aydiner}.
\begin{figure}[htbp!]
\centering
\includegraphics[scale=0.6]{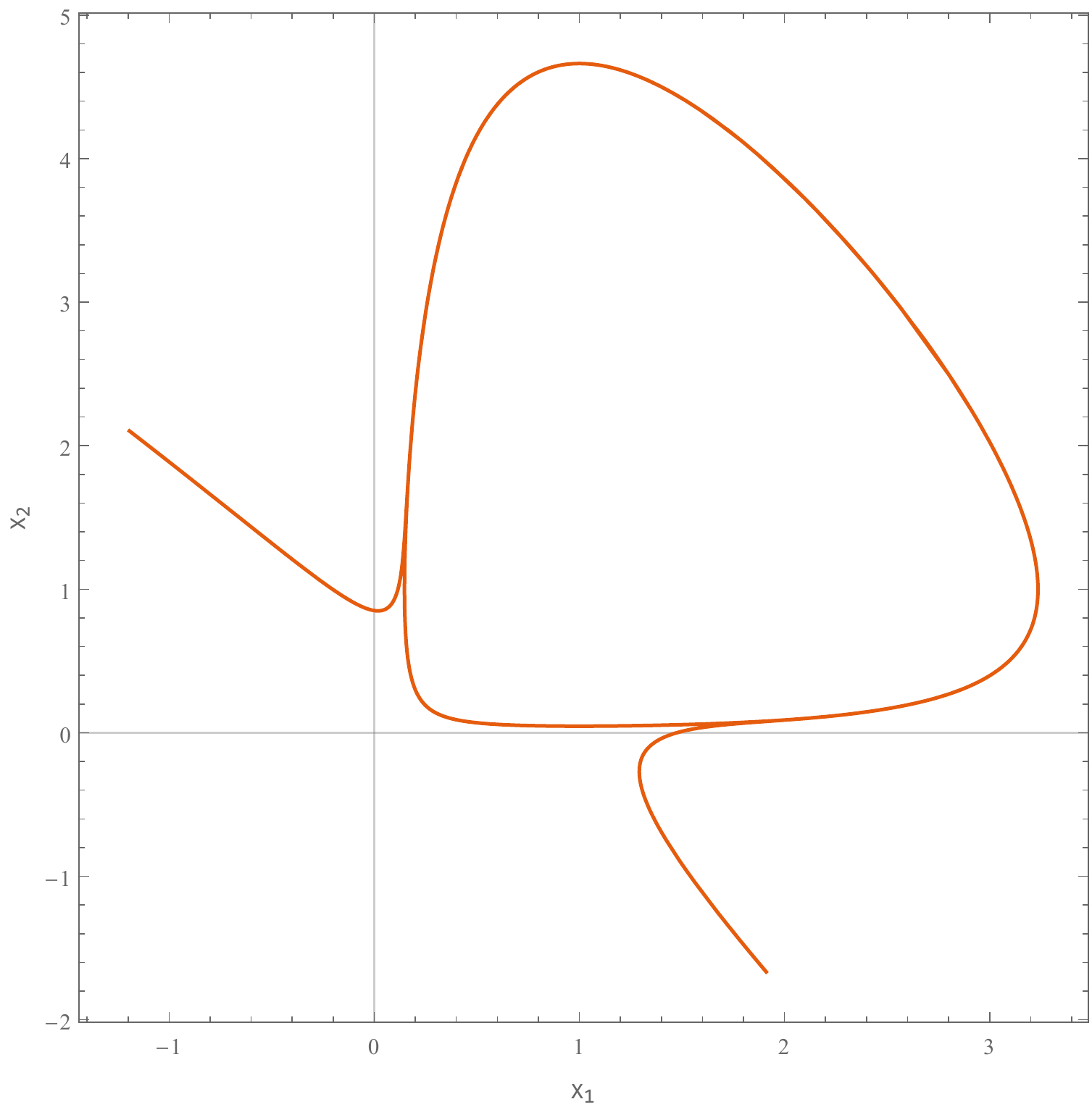}
\caption{Dark matter and dark energy interaction in terms of the dimensionless parameters $x_{2}$ and $x_{1}$, respectively.} 
\label{fig:cycle}
\end{figure}
\\

In Fig. (\ref{fig:cycle}) we show the numerical solution of the Eqs. (\ref{eq:contx1}) and (\ref{eq:contx2}) with the conditions $r_{1} < 0$, $r_{2} >0$ and $\beta(z) > 0$. As can be observed, a cyclic behavior between the dark matter and dark energy can be obtained, but, as the system evolves to the future both components exhibit a divergent behavior, therefore, contrary to the results obtained in \cite{aydiner}, in our case the cyclic behavior is only a transient stage. This is characteristic of universes with future singularity, in general all physical invariants diverge \cite{odint}. From this solution we can infer that at some phase the energy transference between the dark matter and dark energy is equal in both directions but {\it close} to the future singularity this is not longer true. Additionally, also from this solution can be seen that the system is globally unstable. If we consider the case $\beta(z) < 0$, i.e., we change the sign of the curvature parameter, both components have a growing behavior at all stages and no cyclic nature is obtained.
    	   
\section{Final remarks}
\label{sec:final}
Under the scheme of interacting fluids and a holographic cut-off for the dark energy density, we study the resulting dynamics induced by the interaction $Q$-term constructed previously for a FLRW curved spacetime. For this interacting model the presence of a future singularity is allowed.\\

The emerging dynamics can be written as a set of equations which have the structure of the Lotka-Volterra equations for different interacting {\it species}, but, instead constant coefficients we have some functions of the redshift parameter. The virtue of the model used in this work relies on the interaction $Q$-term. As commonly done in the literature, the $Q$-term is given a priori by some different Ansatzes, however in our description we use the form of the interaction term that was constructed by considering a holographic cut-off for the dark energy density given by the Hubble scale together with a Chevallier-Polarsky-Linder type parametrization for the coincidence parameter. This parametrization allows to induce a Type III future singularity in the model. As was demonstrated in this work, this interaction $Q$-term can be written as the product of the dark energy and dark matter densities, this is an important step in order to obtain the arrangement of the Lotka-Volterra type equations. It is important to point out that the flat case can not be obtained from this construction since the future singularity can be obtained only for a curved spacetime under this holographic description.\\

For a $Q$-term given by the product of the densities, the energy transference among the dark energy and dark matter it is equal in both directions (this also depends on the sign of the interaction term), this can be visualized as a cyclic behavior between both components, however, in our description this cyclic behavior is valid only at some phase of the cosmic evolution since the model progress towards a future singularity. As shown, near the singularity, the associated densities to the components of the universe diverge. This suggests that only near the singularity the physical quantities become large \cite{odn}. This differs from the results obtained in Ref. \cite{aydiner}, where the evolution tends to a chaotic attractor. Then, from the outcome obtained in this work, we can argue that the model is globally unstable in the sense of dynamical system, i.e., no bounded trajectories will be obtained in the phase space description.\\

Finally, one way to strengthen the results obtained in this work is given by comparing the model with observational data. Recent results show that some dark energy models in which spatial curvature is included, are not discarded by observations \cite{fin1, fin2}. On the other hand, there are certain advances that allow to establish that the interaction in the dark sector could exist, however, the results are not conclusive since the form of the $Q$-term determines the nature of the interaction \cite{q5}. We intend to return to this point in the near future by considering the interaction term obtained in this work.      

\section*{Acknowledgments}
M.C. work has been supported by S.N.I. (CONACyT-M\'exico). G. M. N. also acknowledges support from a CONACyT scholarship and DGRI-DGUEP-UV.

\end{document}